\documentclass[12pt]{article}
\usepackage{graphicx}
\usepackage{epsfig}
\usepackage{appendix}
\begin{document}
\title{Bounds on the slope and curvature of Isgur-Wise function in a QCD inspired quark model}
\author{$^{1}$Bhaskar Jyoti Hazarika and $^{2}$D K Choudhury\\
$^{1}$Dept of Physics,Pandu College,\\
Guwahati-781012,India\\
$^{2}$Dept. of Physics, Gauhati University, Guwahati-781014,India}
\date{}
\maketitle
\begin{abstract}
 The QCD inspired potential  model persued by us earlier has been recently modified to incorporate an additional factor $'c'$ in the  linear cum Coulomb potential. While it felicitates the inclusion of standard confinement parameter $b=0.183 GeV^{2}$, unlike in previous work, it still falls short of explaining the Isgur-Wise function for the $B$ mesons without adhoc adjustment of the strong coupling constant .\\
In this  work, we determine the factor $'c'$  from the experimental values of decay constants and masses and  show that the reality constraint on $'c'$ yeilds bounds on the strong coupling constant  as well on slope and curvature of Isgur-Wise function  allowing more flexibility to the model. \\

\end{abstract}
\section{Introduction}
In recent years,considerable experimental and theoretical efforts have been undertaken to understand the physics of hadrons containing a heavy quark \cite{1}.The Isgur-Wise function \cite{2} is an important quantity in this area of hadron physics.It is in this spirit, that this function has been studied in various quark models \cite{3}---\cite{12} besides QCD sum rule approach \cite{13},the MIT bag model \cite{14} and the Skryme model \cite{15}.\\
Since one of the basic ingredients of the IW function is the hadron wavefunction involving heavy quark \cite{3}--\cite{12}, it is therefore meaningful to test any specific QCD inspired quark model by calculating the IW function and study it phenomenologically.\\
Sometimes back, a specific QCD inspired quark  model was proposed by us \cite{16} which had later been used to calculate IW function as well \cite{17,18,19}.\\
One of drawback of the model is that significant confinement effect could not be accomodated in the model\cite{16,17,18} due to perturbative constraints coming from using the Dalgarno's method \cite{20}.Only recently \cite{19}, standard confinement effect $b=0.183 GeV^{2}$ \cite{21} was accomodated in the   improved version of QCD inspired quark model brought through the introduction of parameter $'c'$ in the potential : $ V=\frac{-4\alpha_{s}}{3r}+br+c $   taking $c$ $\sim $   $1GeV$ as its natural scale fixing and  $ A_{0}$ = $1$ where $ A_{0}$ is an undetermined factor appearing in the series solution of the Schr\"{o}dinger equation [ Eq.(8) of Ref.19].In earlier works\cite{16,17,18}, the unknown coefficient $cA_{0}$ occured in the wavefunction was set to zero.\\
One of the drawback of work [19]was the adhoc enhancement of strong coupling constant was needed to take into account of the slope and curvature of $B$,$B_{s}$ and $B_{c}$ mesons.\\

In this work, we take an alternative strategy to remove this adhoc enhancement.We use the wavefunction at the origin involving the unknown coefficient $cA_{0}$ and fix it from the experimental values of masses and decay constants directly.The reality constraint on $cA_{0}$ will then yeild lower bounds on the strong coupling  constant $\alpha_{s}$, which would lead to the upper bounds on the slope and curvature of the IW function.\\                                       
 The rest of the paper is organised as follows : Section 2 contains the theory of the improved QCD inspired quark model, Section 3 encloses the results  and in  Section 4 we draw conclusion and remarks.\\

\section{Theory} 
\subsection{The Wavefunction}
    The spin independent Fermi-Breit Hamiltonian for ground state ($l=0$),neglecting the contact term proportional to  $\delta^{3}$  is \cite{16,17} :                      \begin{eqnarray}                                                                H \nonumber&=&H_{o}+H^{\prime}\\&=&-\frac{\nabla^{2}}{2\mu}-\frac{4\alpha_s}{3r}+br+c                                                                         \end{eqnarray}
where $\alpha_{s}$ is the running coupling constant , $b$ is the confinememt parameter and $c$ is another parameter whose significance will be cleared later.\\
In this work,the $\alpha_{s}$ values are taken from the V -scheme \cite{18,26,27} as done in \cite{19} which are large as compared to those of $\overline{MS}$-scheme.It is necessary as large $\alpha_{s}$ values lead to better results for slope and curvature of Isgur-Wise function in earlier work \cite{17,18,19}.\\
In earlier work \cite{17,18} , large confinement was found to be inconvenient in the calculation of slope and curvature of Isgur -Wise function which was a limitation of the model.So, our aim has been focussed in the inclusion of larger confinement and hence we retain the same choice of $b=0.183 GeV^{2}$ \cite{19,21} to investigate whether this approach leads to better results or not.It is worth notable that inclusion of $c$ in the analysis \cite{19} allows larger $b$ useful , so we don't think another choice of $b$. \\ 
  With ,$H_{o}=-\frac{\nabla^{2}}{2\mu}-\frac{4\alpha_{s}}{3r}$ as the parent Hamiltonian and $H^{\prime}=br+c$ as the perturbed Hamiltonian , we obtain a ground state wavefunction upto the first order correction using the Dalgarno method \cite{20} of stationary state perturbation theory as :\\ 

\begin{equation}
\psi_{conf}\left(r\right)= N\left(cA_{0}+\frac{1}{\sqrt{\pi a_{0}^{3}}}-\frac{\mu ba_{0}r^{2}}{\sqrt{\pi a_{0}^{3}}}\right)e^{-\frac{r}{a_{0}}}
\end{equation}
 where $A_{0}$ is the unknown coefficient appearing in the series solution of the Dalgarno method \\

Including the relativistic effect \cite{22,23}, the wavefunction is  :\\
 
\begin{equation}
\psi_{conf+rel}\left(r\right)= N^{\prime}\left(cA_{0}+\frac{1}{\sqrt{\pi a_{0}^{3}}}-\frac{\mu ba_{0}r^{2}}{\sqrt{\pi a_{0}^{3}}}\right)\left(\frac{r}{a_{0}}\right)^{-\epsilon} e^{-\frac{r}{a_{0}}}
\end{equation}

Here $a_{0}$ is given by:\\

\begin{equation}
a_{0}=\frac{3}{4\mu \alpha_{s}}
\end{equation}

and\\

\begin{equation}
\epsilon=1-\sqrt{1-\frac{4\alpha_{s}}{3}}
\end{equation}

$N$ and $N^{\prime}$ are the normalization constants given by :\\

\begin{equation}
N^{2}=\frac{1}{1+\frac{45\mu^{2}b^{2}a_{0}^{6}}{8}-3\mu ba_{0}^{3}+\pi a_{0}^{3}c^{2}A_{0}^{2}+\frac{2cA_{0}\pi a_{0}^{3}}{\sqrt{\pi a_{0}^{3}}}-\frac{3\pi a_{0}^{6}cA_{0}\mu b}{\sqrt{\pi a_{0}^{3}}}}
\end{equation}
and 
\begin{equation}
N^{\prime^{2}}=\frac{2^{7-2\epsilon}}{\Gamma\left(3-2\epsilon\right)X_{1}}
\end{equation}
where $X_{1}$ is given in APPENDIX A.\\

We note that the equations (2),(3),(6)and (7) are obtained from Eq.(4),(6),(5) and (7) of Ref.[19] exhibiting explicit dependence of $cA_{0}$ in them.
\subsection{Fixing of the coefficient  $ cA_{0}$}
 The wavefunction at the origin (WFO), is related to the decay constant $f_{p}$ and the mass of the pseudoscalar meson $M_{p}$  through the relation \cite{16,24}:\\

\begin{equation}
|\psi\left(0\right)|^{2}=\frac{f_{p}^{2}M_{p}}{12}
\end{equation}

Again  from equation (2), we have :\\

\begin{equation}
 |\psi\left(0\right)|^{2}=N^{2}[c^{2}A_{0}^{2}+\frac{1}{\pi a_{0}^{3}}+\frac{2cA_{0}}{\sqrt{\pi a_{0}^{3}}}]
\end{equation}
Using equation(6), we arrive at the quadratic equation for $cA_{0}$:
\begin{equation}
A^{\prime}\left(cA_{0}\right)^{2}+B^{\prime}\left(cA_{0}\right)+C^{\prime}=0
\end{equation}
where\\
\begin{equation}
A^{\prime}= \pi a_{0}^{3}|\psi\left(0\right)|^{2}-1
\end{equation}
\begin{equation}
B^{\prime}=2\sqrt{\pi a_{0}^{3}}|\psi\left(0\right)|^{2}-3\mu b a_{0}^{3}\sqrt{\pi a_{0}^{3}}|\psi\left(0\right)|^{2}
\end{equation}
and
\begin{equation}
C^{\prime}=|\psi\left(0\right)|^{2}\left[1+\frac{45 \mu^{2} b^{2}a_{0}^{6}}{8}-3\mu b a_{0}^{3}\right]-\frac{1}{\pi a_{0}^{3}}
\end{equation}
Using the experimental values of $f_{p}$ and $M_{p}$ \cite{25} , we determine $|\psi\left(0\right)|^{2}$ from equation(9) which in turn will yeild two solutions for $cA_{0}$ in  equation (10):
\begin{equation}
cA_{0}=\frac{-B^{\prime}\pm \sqrt{B^{\prime^2}-4A^{\prime}C^{\prime}}}{2A^{\prime}}
\end{equation}
which will depend on $\mu$,$M_{P}$,$f_{P}$ and $\alpha_{s}$.The solution corresponding to the +ve(-ve) sign of equation(15)will be termed as +ve(-ve) solution hereafter.It will be shown numerically that for a given $\mu$, $M_{P}$,and $f_{P}$ ,$\alpha_{s}$ reaches the minimum value when the following condition is satisfied :
\begin{equation}
B^{\prime^2}-4A^{\prime}C^{\prime}=0
\end{equation}
The formalism involving Eq.(5)-(16) is strictly valid only without relativistic effect as the wavefunction at the origin with such effect [Eq.(3)] is not well defined due to its singularity at the origin.For a subsequent analysis ,we assume that $cA_{0}$ does not deviate significantly from its non-relativistic value so that it can be used to calculate the slope and curvature of the IW function even without relativistic effect.\\

\subsection{Charge radius (slope)and convexity parameter (curvature) of I-W function}

The Isgur-Wise function is written as \cite{2,17} :
\begin{eqnarray}
\xi\left(v_{\mu}.v^{\prime}_{\mu}\right)\nonumber&=&\xi\left(y\right)\\&=&1-\rho^{2}\left(y-1\right)+ C\left(y-1\right)^{2}+...
\end{eqnarray}
where 
\begin{equation}
y= v_{\mu}.v^{\prime}_{\mu}
\end{equation}
and $v_{\mu}$ and $v^{\prime}_{\mu}$  being the four velocity of the heavy meson before and after the decay.The quantity $\rho^{2}$  is the slope of I-W function at $y=1$ and known as charge radius :\\

\begin{equation}
\rho^{2}= \left. \frac{\partial \xi}{\partial y}\right.|_{y=1}
\end{equation}
The second order derivative is the curvature of the I-W function known as convexity parameter :\\

\begin{equation}
C=\frac{1}{2}\left[\frac{\partial^2 \xi}{\partial^2 y}|_{y=1}\right]
\end{equation}
For the heavy-light flavor mesons the I-W function can also be written as \cite{6,17} :

\begin{equation}
\xi\left(y\right)=\int_{0}^{+\infty} 4\pi r^{2}\left|\psi\left(r\right)\right|^{2}\cos pr dr
\end{equation}
where
\begin{equation}
p^{2}=2\mu\left(y-1\right)
\end{equation}
Equation (21) holds good for both relativistic and nonrelativistic case.The wavefunction $\psi\left(r\right)$ takes different form for both the cases. Without relativistic effect, it is given by equation(2) and with relativistic effect it is given by (3).\\
  
With the wavefunction(2)in equation(10) i.e. including confinement only the charge radius $\rho_{conf}^{2}$ and convexity parameter $C_{conf}$ are respectively given by:\\
\
\begin{equation}
\rho_{conf}^{2}=\frac{\mu^{2}[24\pi c^{2}A_{0}^{2}a_{0}^{5}+24a_{0}^{2}+630\mu^{2}b^{2}a_{0}^{8}+48cA_{0}\sqrt{\pi a_{0}^{7}}-180cA_{0}\mu b\sqrt{\pi a_{0}^{13}}-180\mu ba_{0}^{5}]}{8\pi c^{2}A_{0}^{2}a_{0}^{3}+8+45\mu^{2}b^{2}a_{0}^{6}+16cA_{0}\sqrt{\pi a_{0}^{3}}-24\mu bcA_{0}\sqrt{\pi a_{0}^{3}}-24\mu ba_{0}^{3}}
\end{equation}

and  :
\begin{equation}
C_{conf}=\frac{\mu^{4}[60\pi c^{2}A_{0}^{2}a_{0}^{7}+60a_{0}^{4}+4725\mu^{2}b^{2}a_{0}^{10}+120cA_{0}\sqrt{\pi a_{0}^{10}}-840cA_{0}\mu b\sqrt{\pi a_{0}^{17}}-840\mu ba_{0}^{7}]}{16\pi c^{2}A_{0}^{2}a_{0}^{3}+16+90\mu^{2}b^{2}a_{0}^{6}+32cA_{0}\sqrt{\pi a_{0}^{3}}-48\mu bcA_{0}\sqrt{\pi a_{0}^{3}}-48\mu ba_{0}^{3}}
\end{equation}
With the wavefunction (3)in equation (10) i.e. including both relavistic and confinement effect the charge radius $\rho_{conf+rel}^{2}$ and convexity parameter $C_{conf+rel}$ are given by :\\

\begin{equation}
\rho_{conf+rel}^{2}=\frac{\mu^{2}a_{0}^{2}\left(4-2\epsilon\right)\left(3-2\epsilon\right)[X_{1}]}{4[X_{2}]}
\end{equation}
and
\begin{equation}
C_{conf+rel}=\frac{\mu^{4}a_{0}^{4}\left(6-2\epsilon\right)\left(5-2\epsilon\right)\left(4-2\epsilon\right)\left(3-2\epsilon\right)[X_{3}]}{96[X_{2}]}
\end{equation}
where $X_{1}$,$X_{2}$  and $X_{3}$  are given in Appendix.\\

We note that equations (25) and (26) are equivalent to equations (18) and (19)of Ref[19] exhibiting explicit $cA_{0}$ dependence.

\section{Results}

\subsection{Values of $cA_{0}$ and lower bounds on $\alpha_{s}$}
As noted earlier , $cA_{0}$ depends on $\mu$,$M_{P}$,$f_{P}$ and $\alpha_{s}$.In fig.1(a-e) we plot $cA_{0}$ vs $\alpha_{s}$ for $D$,$D_{s}$,$B$,$B_{s}$ and $B_{c}$ mesons .It shows that $\alpha_{s}$ tends to reach  the minimum value when two solutions of Eq.(11) almost merge satisfying the condition(16).This feature is true for any set of the parameters $\mu$, $f_{p}$ and $M_{p}$. In table 1 ,we give the lower bounds on $\alpha_{s}$ for mesons having $c$ and $b$ quarks.\\

The dependence of $cA_{0}$ on $\alpha_{s}$ and $\mu$ can be noted as follows :\\
With constant  $\mu$ , $cA_{0}$ decreases  with $\alpha_{s}$ values rising and vice-versa.On the other hand, with constant $\alpha_{s}$ , $cA_{0}$ increases(decreases) with increase (decrease) in $\mu$.\\

\begin{figure}
\includegraphics[width=4.5in,angle=270]{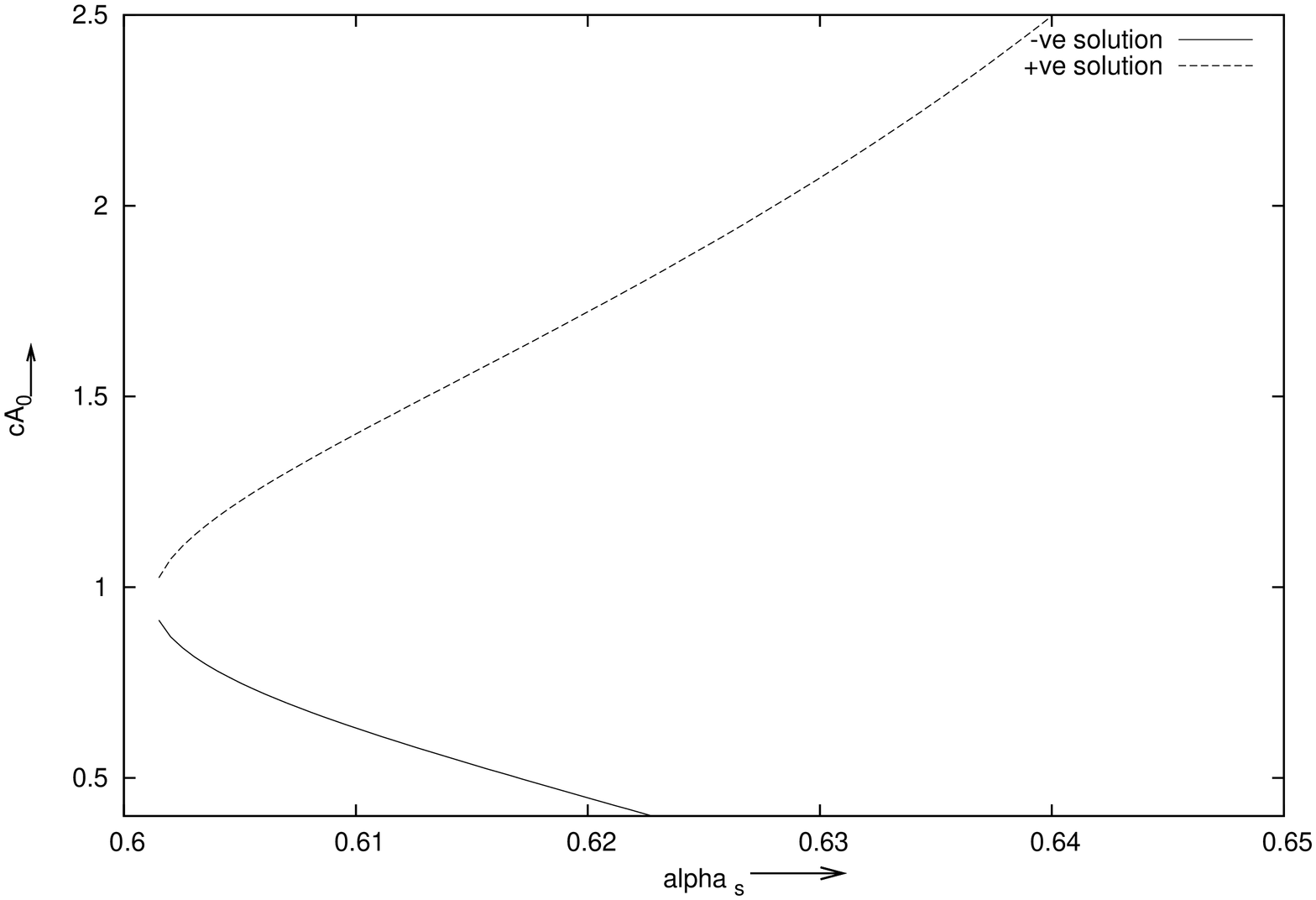}
\caption{Variation of $cA_{0}$ vs $\alpha_{s}$ for $D$ Meson.The +ve (-ve) solution of Eq.15 corresponds to the dashed (solid) line and the two lines nearly coincide at $\alpha_{s}$  $\sim $   $0.601$ ,  the lower bound on $\alpha_{s}$ corresponding to the solution of Eq.16 for D Meson. }
\end{figure}

\begin{figure}
\includegraphics[width=4.5in,angle=270]{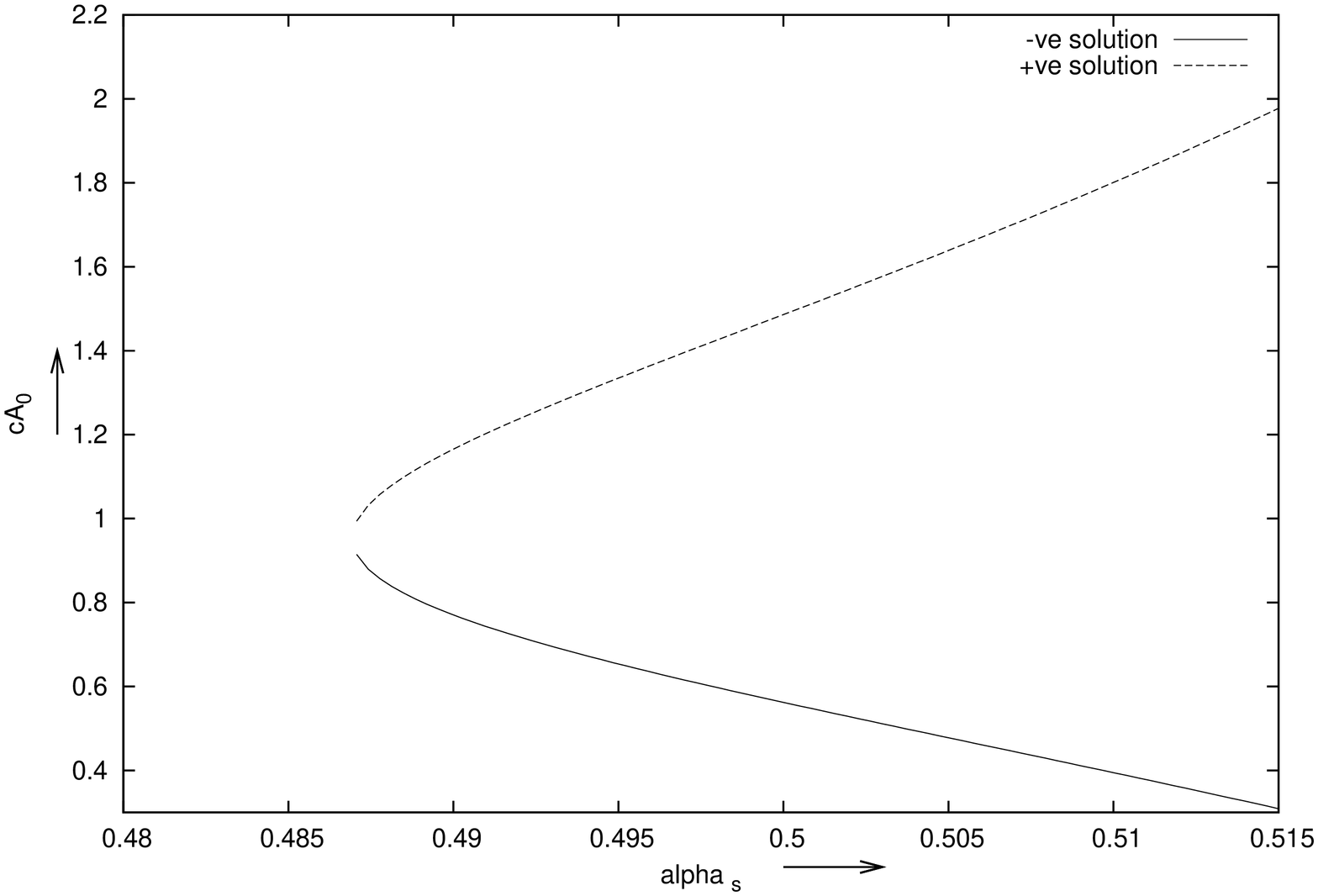}
\caption{Variation of $cA_{0}$ vs $\alpha_{s}$ for $D_{s}$ Meson.The +ve (-ve) solution of Eq.15 corresponds to the dashed (solid) line and the two lines nearly coincide at $\alpha_{s}$ $\sim $   $0.49$ ,  the lower bound on $\alpha_{s}$ corresponding to the solution of Eq.16 for $D_{s}$ Meson.}
\end{figure}

\begin{figure}
\includegraphics[width=4.5in,angle=270]{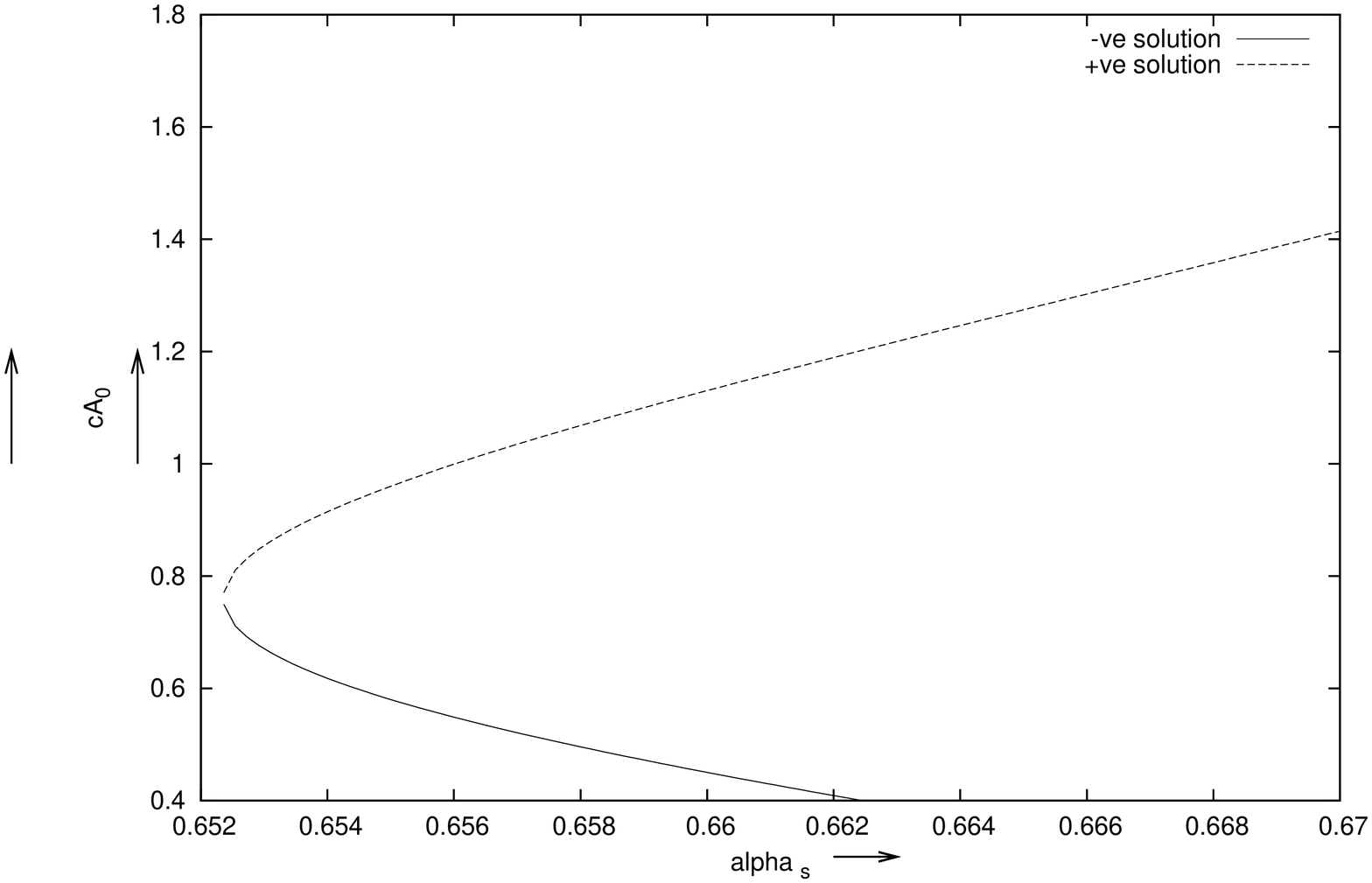}
\caption{Variation of $cA_{0}$ vs $\alpha_{s}$ for $B$ Meson.The +ve (-ve) solution of Eq.15 corresponds to the dashed (solid) line and the two lines nearly coincide at $\alpha_{s}$ $\sim $   $0.652$ ,  the lower bound on $\alpha_{s}$ corresponding to the solution of Eq.16 for $B$ Meson.}
\end{figure}

\begin{figure}
\includegraphics[width=4.5in,angle=270]{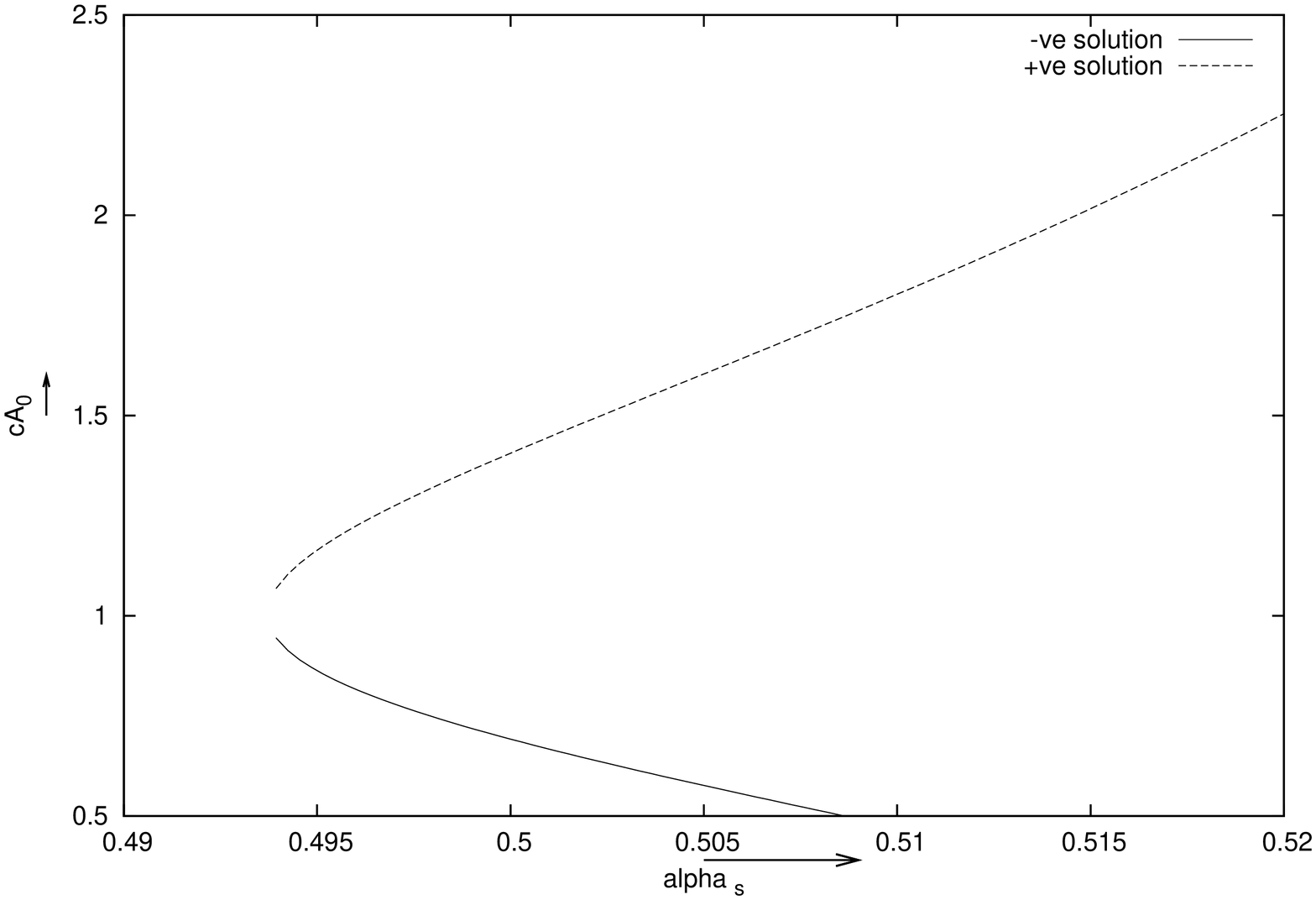}
\caption{Variation of $cA_{0}$ vs $\alpha_{s}$ for $B_{s}$ Meson.The +ve (-ve) solution of Eq.15 corresponds to the dashed (solid) line and the two lines nearly coincide at $\alpha_{s}$ $\sim $   $0.493$ ,  the lower bound on $\alpha_{s}$ corresponding to the solution of Eq.16 for $B_{s}$ Meson.}
\end{figure}

\begin{figure}
\includegraphics[width=4.5in,angle=270]{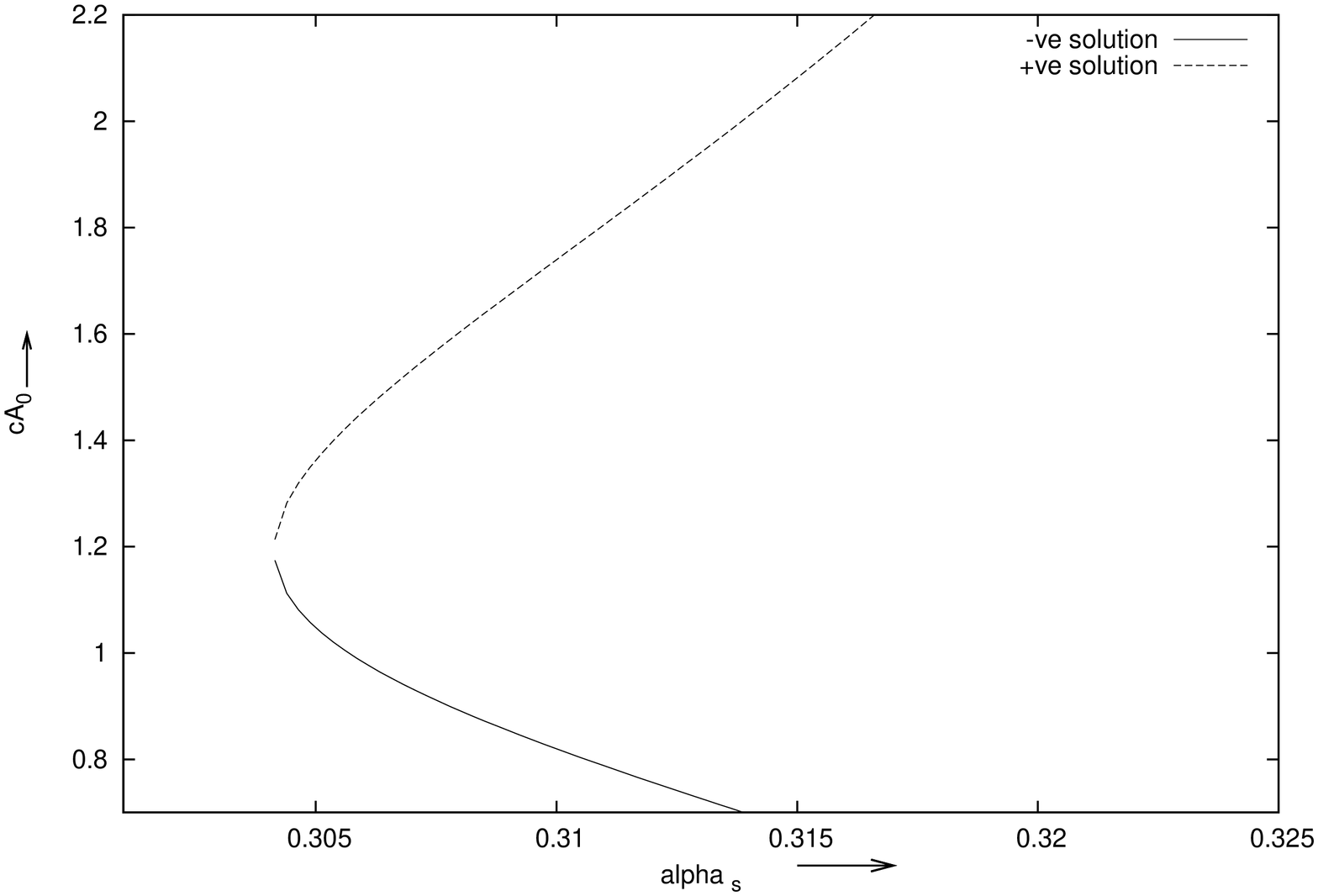}
\caption{Variation of $cA_{0}$ vs $\alpha_{s}$ for $B_{c}$ Meson.The +ve (-ve) solution of Eq.15 corresponds to the dashed (solid) line and the two lines nearly coincide at $\alpha_{s}$ $\sim $   $0.302$ ,  the lower bound on $\alpha_{s}$ corresponding to the solution of Eq.16 for $B_{c}$ Meson.}
\end{figure}

\begin{table}
\begin{center}
\caption{ Lower Bounds on  $\alpha_{s}$ }
\begin{tabular}{|c|c|c|c|c|c|c| }\hline
Mesons& Quark content&$\mu$(GeV)&$M_{p}$(GeV)&$f_{p}$(GeV)&$cA_{0}$& Lower bound \\
 & &Ref[25] &Ref[25] &Ref[25]& &on $\alpha_{s}$\\\hline
$D$&$c\bar{u}/c\bar{d}$&0.276&1.869&0.192&0.9665&$\sim $ $0.601$\\\hline
$B$&$\bar{b}u/\bar{b}d$&0.315&1.968&0.157&0.7653&$\sim $ $0.652$\\\hline
$D_{s}$&$c\bar{s}$&0.368&5.279&0.210&1.1967&$\sim $ $0.49$\\\hline
$B_{s}$&$\bar{b}s$&0.44&5.279&0.171&0.999&$\sim $ $0.493$\\\hline
$B_{c}$&$\bar{b}c$&1.18&5.37&0.36&1.167&$\sim $ $0.302$\\\hline
\end{tabular}
\end{center}
\end{table}

\subsection{Bounds on slope and curvature of the IW function}
Using the lower bounds on $\alpha_{s}$ for each heavy-light and heavy-heavy mesons,we obtain upper bounds on the slope and curvature of the I-W function  using equations (23),(24),(25) and (26). They are listed in table 2.We note that with increasing $\alpha_{s}$ values, the slope and curvature decreases and henceforth the lower bound on $\alpha_{s}$ corresponds to the upper bound on $\rho^{2}$ and $C$.\\

In table 3,we record the predictions of the slope and curvature of the IW function in various models while in table 4,we reproduce the corresponding predictions of the model of Ref.[19] with $c=1 GeV$ and $A_{0}=1$ in V-scheme \cite{26,27} for various mesons.Two values for $B$, $B_{s}$ and $B_{c}$ mesons are shown where case- a) represents the actual values for $\rho^{2}$ and $C$ in that work with $\alpha_{s}=0.261$;while case-b) represents those for adhoc adjustable value of $\alpha_{s}=0.60$ in order to show the usefulness of large $\alpha_{s}$ as mentioned in Ref 19.The  $\alpha_{s}$ are already large for $D$ and $D_{s}$ mesons,so no two values are shown. \\

\begin{table}
\begin{center}
\caption{ Upper Bounds on Slope and Curvature }
\begin{tabular}{|c|c|c|c|c|}\hline
Meson&\multicolumn{2}{c|}{Slope $\rho^{2}$}&\multicolumn{2}{c|}{Curvature $C$}\\
\cline{2-5}
(Quark &\multicolumn{1}{l|}{Without relati-}&With relati- &\multicolumn{1}{l|}{Without relati-}&With relativi-\\
Content)&\multicolumn{1}{l|}{vistic effect}&vistic effect&\multicolumn{1}{l|}{vistic effect}&vistic effect\\\hline
$D(c\bar{u}/c\bar{d})$&\multicolumn{1}{l|}{6.78}&1.675&\multicolumn{1}{l|}{13.19}&5.138\\\hline
$B(\bar{b}u/\bar{b}d)$&\multicolumn{1}{l|}{5.78}&1.016&\multicolumn{1}{l|}{9.58}&1.29\\\hline
$D_{s}(c\bar{s})$&\multicolumn{1}{l|}{9.115}&3.067&\multicolumn{1}{l|}{26.48}&14.32\\\hline
$B_{s}(\bar{b}s)$&\multicolumn{1}{l|}{11.92}&2.652&\multicolumn{1}{l|}{34.49}&6.902\\\hline
$B_{c}(\bar{b}c)$&\multicolumn{1}{l|}{28.46}&10.39&\multicolumn{1}{l|}{219.46}&45.23\\\hline
\end{tabular}
\end{center}
\end{table}
\begin{table}
\begin{center}
\caption{Predictions of the slope and curvature of the IW function in various models.}
\begin{tabular}{|c|c|c|}\hline
Model& Value of $\rho^{2}$ &Value of curvature $C$\\\hline
Yaouanc et al \cite{28}&$\ge 0.75$&..\\
Yaouanc et al \cite{12}&$\ge 0.75$&$\ge 0.47$\\ 
Rosner et al \cite{29}&1.66&2.76\\
Mannel et al\cite{30,31}&0.98&0.98\\
Pole Ansatz \cite{32}&1.42&2.71\\
MIT Bag Model \cite{14}&2.35&3.95\\
Simple Quark Model \cite{3}&1&1.11\\
Skryme Model \cite{15}&1.3&0.85\\
QCD Sum Rule \cite{13}&0.65&0.47\\
Relativistic Three Quark Model \cite{4}&1.35&1.75\\
Infinite Momentum Frame Quark Model \cite{5}&3.04&6.81\\\hline
\end{tabular}
\end{center}
\end{table}
\begin{table}
\begin{center}
\caption{Predictions of the slope and curvature of the I-W function in the QCD inspired quark model according to Ref[19] with $c=1$ and $A_{0}=1$ taking relavistic and confinement effect in V-scheme.This table is nothing , but the copy of the last rows of tables 1,2,3 of Ref 19 .}
\begin{tabular}{|c|c|c|c|}\hline
Meson&$\alpha_{s}$&slope ($\rho^{2}$)&curvature($C$)\\\hline
$D$&0.625&1.136&5.377\\\hline
$D_{s}$&0.625&1.083&3.583\\\hline
$B$&$a)$0.261&$a)$128.128&$a)$5212\\
   &$b)$0.60&$b)$1.329&$b)$7.2\\\hline
$B_{s}$&$a)$0.261&$a)$112.759&$a)$4841\\
  &$b)$0.60&$b)$1.257&$b)$4.379\\\hline
$B_{c}$&$a)$0.261&$a)$44.479&$a)$2318\\
     &$b)$0.60&$b)$1.523&$b)$0.432\\\hline
\end{tabular}
\end{center}
\end{table}

\section{Conclusion and Remarks}

In this paper,we have shown that the reality bound on $cA_{0}$  puts lower limit on  $\alpha_{s}$ and correspondingly upper limit on $\rho^{2}$ and $C$. \\

Furthermore,with $cA_{0}$ , the upper bounds on  $\rho^{2}$ and $C$ decrease which is evident from the above list of bounds[table-2]. The estimated upper bounds on $\rho^{2}$ and $C$ for all the mesons are found to be consistent with  other models and data [table-3] without making any adhoc enhancement of the strong coupling constant as had been done in ref (19)[table-4] .From the phenomenological point of view we note that in the nonrelativistic limit ,the universal form factor and Isgur-Wise function for semileptonic decay $B \rightarrow {D^{*}l\nu}$ are identical when subleading terms in velocity and terms of order $ O\left(\frac{E_{b}}{m_{Q}}\right)$ are neglected with $E_{b}$ as the binding energy and $m_{Q}$ as the mass of heavy quark \cite{33}.However even if we make calculation for the universal form factor for finite mass, we obtain to first order in $\left(y-1\right)$ as $0.8$ -$2.57\left(y-1\right)$ which seems to be satisfactory \cite{33,34}.\\

It is worth notable that in the limit $cA_{0}\rightarrow 0$, there will be no bounds on $\alpha_{s}$ as well as on $\rho^{2}$ and $C$ ; rather fixed values of $\alpha_{s}$ have to be used to get definite set of $\rho^{2}$ and $C$.So,in that case, the analysis will turn to  that of ref [17,18] where large confinement could not be (i.e.$b=0.183 GeV^{2}$) incorporated e.g. tables -(1,3) of ref[17] and tables -(2,3) of ref[18].\\
  
We conclude this paper with a comment on the physical significance of the factor $'c'$ that has become so crucial for our analysis of bounds on slope and curvature.\\
It is common wisdom that a constant potential like $'c'$ just scales the energies and doesnot affect the wavefunction nor does it change physics.This can be seen from the hydrogen atom problem with the potential $ V(r)= =\frac{-A}{r}+c $.However , if one uses $'c'$ as the perturbation instead of as parent in the Dalgarno method of perturbation theory \cite{20},the wavefunction for the $H$-atom becomes:\\
$\psi\left(r\right)= N_{1}\left(cA_{0}+\frac{1}{\sqrt{\pi a_{0}^{3}}}\right)e^{-\frac{r}{a_{0}}}$\\
to be compared with the wavefunction with $'c'$ as parent:\\
$\psi\left(r\right)=\left(\frac{1}{\sqrt{\pi a_{0}^{3}}}\right)e^{-\frac{r}{a_{0}}}$\\

where the normalization constant:\\

$N_{1}^{2}=\frac{1}{1+\pi a_{0}^{3}c^{2}A_{0}^{2}+\frac{2cA_{0}\pi a_{0}^{3}}{\sqrt{\pi a_{0}^{3}}}}$ \\
Thus, the perturbative child $'c'$ rather than the parent $'c'$ plays the crucial role in the present analysis.\\

\appendix
$X_{1}$, $X_{2}$ and $X_{3}$ are evaluated as : 
\begin{eqnarray}
X_{1}=64\pi c^{2}A_{0}^{2}a_{0}^{3}+64+\mu^{2}b^{2}a_{0}^{6}\left(8-2\epsilon\right)\left(7-2\epsilon\right)\left(6-2\epsilon\right)\left(5-2\epsilon\right)\nonumber\\
+128cA_{0}\sqrt{\pi a_{0}^{3}}-16cA_{0}\mu b\sqrt{\pi a_{0}^{9}}\left(6-2\epsilon\right)\left(5-2\epsilon\right)\nonumber\\
-16\mu ba_{0}^{3}\left(6-2\epsilon\right)\left(5-2\epsilon\right)
\end{eqnarray}

\begin{eqnarray}
X_{2}=64\pi c^{2}A_{0}^{2}a_{0}^{3}+64+\mu^{2}b^{2}a_{0}^{6}\left(6-2\epsilon\right)\left(5-2\epsilon\right)\left(4-2\epsilon\right)\left(3-2\epsilon\right)\nonumber\\
+128cA_{0}\sqrt{\pi a_{0}^{3}}-16cA_{0}\mu b\sqrt{\pi a_{0}^{9}}\left(4-2\epsilon\right)\left(3-2\epsilon\right)\nonumber\\
-16\mu ba_{0}^{3}\left(4-2\epsilon\right)\left(3-2\epsilon\right)
\end{eqnarray}

\begin{eqnarray}
X_{3}=64\pi c^{2}A_{0}^{2}a_{0}^{3}+64+\mu^{2}b^{2}a_{0}^{6}\left(10-2\epsilon\right)\left(9-2\epsilon\right)\left(8-2\epsilon\right)\left(7-2\epsilon\right)\nonumber\\
+128cA_{0}\sqrt{\pi a_{0}^{3}}-16cA_{0}\mu b\sqrt{\pi a_{0}^{9}}\left(8-2\epsilon\right)\left(7-2\epsilon\right)\nonumber\\
-16\mu ba_{0}^{3}\left(8-2\epsilon\right)\left(7-2\epsilon\right)
\end{eqnarray}
Not only the above expressions ,but all the integrals in the analysis are evaluated with the help of Gamma function given by :
\begin{equation}
\frac{\Gamma\left(n+1\right)}{\alpha^{n+1}}=\int_{0}^{+\infty}  r^{n}e^{-\alpha r} dr
\end{equation}

\end{document}